\documentclass[prl,twocolumn,showpacs,amsmath,amssymb,aps]{revtex4}

\usepackage{graphicx}
\usepackage{dcolumn}
\usepackage{amsmath}
\usepackage{bm}

\begin{document}
\title{Role of Bound Magnon in Magnetic Domain Wall Motion}

\author{Tae-Suk Kim$^{1,2}$, J. Ieda$^{1,3}$, and S. Maekawa$^{1,3}$}
\affiliation{$^1$Institute for Materials Research, Tohoku Univ., Sendai 980-8577, Japan  \\
$^2$Department of Physics, Pohang University of Science and Technology, Pohang 790-784, Korea   \\
$^3$CREST, Japan Science and Technology Agency (JST), Tokyo 102-0075, Japan}
\received{\today}

\begin{abstract}
We report on a quantum description of the domain wall (DW) motion under a spin current. 
A bound magnon, which is the zero mode of DW, is found to play a dominant role in DW dynamics. 
The bound magnon acquires its inertia by the hard axis anisotropy and is a free particle even under the spin current. 
The full transfer of spin angular momentum from the spin current to DW via the bound magnon leads to 
the DW motion with the adiabatic velocity, decoupling of spin waves from DW, and no Doppler shift in spin waves. 
\end{abstract}

\pacs{75.45.+j, 75.60.Ch}

\maketitle

 Magnetic DW motion \cite{slonczewski,maekawa} has attracted much interest to both experiments and theories,
owing to its potential device applications such as DW logic \cite{DW_logic} and DW memory \cite{DW_memory}.
DW motion under magnetic fields is now well documented and
understood \cite{slonczewski} in terms of classical Landau-Lifshitz-Gilbert (LLG) equations.
Recently many experimental groups \cite{exp_DW1,exp_DW2,exp_DW3,exp_DW4,exp_DW5,exp_DW6} 
have observed the magnetic DW motion under spin currents.  
However, theoretical description \cite{bazaliy,tatara,barnes,li_zhang,thiaville}
is very controversial over several issues. 
Even for a perfect ferromagnetic (FM) nanowire, the hard axis anisotropy was claimed \cite{tatara} 
to induce the intrinsic pinning so that a DW does not move until a finite spin current is applied. 
But this intrinsic pinning is not consistent with translational symmetry of DW \cite{barnes}.
In the generalized LLG equation approach to the DW motion,
the spin current gives rise to the so-called nonadiabatic \cite{barnes,li_zhang,thiaville}
as well as adiabatic torques. 
The relative magnitude of nonadiabatic and Gilbert damping torques remains an unresolved problem.

 In this paper we study the DW motion under spin currents, using the one dimensional $s$-$d$ model
Hamiltonian and the full quantum mechanical description of the DW motion. 
In the LLG equation approach, macroscopic magnetization is treated as a classical vector
with fixed magnitude and two Euler angles. In this work we treat both electron and local spin systems 
quantum mechanically  in order to 
address a microscopic mechanism for spin transfer from a spin current to a DW.    
We find that a bound magnon (with zero energy) in the DW plays a dominant role 
in absorbing angular momentum from the spin current and thereby in the DW motion. 
This bound magnon is the zero mode of magnetic soliton (DW) \cite{soliton}. 
The mass of the bound magnon is derived quantum mechanically for the first time
and is in agreement with the classical DW or D\"{o}ring mass \cite{doring}. 
Our study shows that the bound magnon is a free particle even under a spin current  
and a DW can start to move under any finite spin currents for a perfect FM nanowire.
The bound magnon-mediated spin transfer mechanism is compatible with the balance of 
 the nonadiabatic and Gilbert damping torques. 
The role of the bound magnon is very similar to that of zero phonon mode in solitons of charge density wave 
state \cite{cdw}.

  We consider the one-dimensional $s$-$d$ model to study the DW motion which is driven by 
spin currents. Our model system consists of three parts: $H = H_e + H_S + H_{eS}$
\begin{eqnarray}
H_e &=& - t \sum_{i\alpha} \left[ c_{i\alpha}^{\dag} c_{i+1\alpha} + H.c. \right]
    - \mu \sum_{i\alpha} c_{i\alpha}^{\dag} c_{i\alpha},  \\
H_S &=& - J \sum_{i} \vec{S}_i \cdot \vec{S}_{i+1}
    - A \sum_i (\vec{S}_i\cdot\hat{z})^2 + K \sum_i (\vec{S}_i \cdot \hat{y})^2,  \\
H_{eS} &=& - J_H \sum_{i} \vec{S}_{ci} \cdot \vec{S}_i.
\end{eqnarray}
Conduction $s$ electrons are described by $H_e$ and are spin polarized by the Hund coupling 
$H_{eS}$ to the ordered local spins. 
$H_S$ describes a system of local spins with an easy $z$-axis ($A>0$) along the wire direction
and a hard $y$-axis ($K>0$). Ferromagnetically ($J>0$) coupled local spins are assumed to have a 
transverse DW. 
 DWs can be induced in the FM wire by ingenious experimental techniques. 
The mutually orthogonal unit vectors, $\hat{x}, \hat{y}$ and $\hat{z}$, define the laboratory frame.

Representing the local spins in terms of two Euler angle fields, $\theta$ and $\phi$, 
the DW structure can be derived by the energy minimization. 
The ordered local spins lie on the easy $x$-$z$ plane ($\phi_w =0, \pi$) 
and are rotated away from the easy axis by the angle 
$\theta_w (z) = 2 \cot^{-1} e^{- (z-q)/\Delta}$. 
$q$ is the DW position, 
$\Delta = \sqrt{Ja^2/2A}$ is the DW width and
$a$ is the lattice spacing between two neighboring spins.
The rotation angle $\theta_w$ at each spin defines the quantization axis along which local spins are aligned.

 Local spins may well fluctuate away from the ordered DW state. Small fluctuating spin fields can be
represented by small fluctuating Euler angle fields. But instead, we adopt in this work the small fluctuating 
transverse spin fields, $S_{ix}$ and $S_{iy}$, which are a more natural description of spin fluctuations.
Using the local coordinate frames defined by the local spin quantization axis, 
the spin fields can be represented in terms of local transverse spin fields,
\begin{eqnarray}
\hat{x} \cdot \vec{S}_i 
  &=& S_{ix} \cos\theta_w(z_i) + S_{iz} \sin\theta_w(z_i),   \\
\hat{y} \cdot \vec{S}_i 
  &=& S_{iy},  \\
\hat{z} \cdot \vec{S}_i 
  &=& S_{iz} \cos\theta_w(z_i) - S_{ix} \sin\theta_w(z_i). 
\end{eqnarray}
Here $S_{iz} = \sqrt{S(S+1) - S_{ix}^2 - S_{iy}^2}$. 
Expanding the transverse spin fluctuations away
from the quantization axis, we find the magnon Hamiltonian $H_{mag} = H_0 + H_1$, 
keeping only up to quadratic terms in transverse spin components. 
\begin{eqnarray}
H_0 &=& \sum_i \sum_{a=x,y} \left[ \frac{J}{2} (S_{ia} - S_{i+1a})^2 
       + A ( \cos^2 \theta_i - \sin^2 \theta_i) S_{ia}^2 \right],  
\end{eqnarray}
and $H_1 = K \sum_i S_{iy}^2$, where $\theta_i = \theta_w(z_i)$. 
The transverse spins in the continuum limit ($S_{ix} - S_{i+1x} \simeq -a \partial S_{x}/\partial z$) 
satisfy the following equations of motion,
\begin{eqnarray}
\label{snmodex}
\hbar \partial_t S_x &=& H_{dw} S_y + 2KS ~ S_y,   \\
\label{snmodey}
\hbar \partial_t S_y &=& - H_{dw} S_x. 
\end{eqnarray}
Here $H_{dw} = - JS a^2 \partial_z^2 + 2AS (\cos^2 \theta_w - \sin^2 \theta_w )$ 
and acts as a Hamiltonian for normal modes of magnons.

Winter \cite{winter} found two types of normal modes for the transverse DW: 
the magnon bound to the DW and the extended spin waves. 
The static Schr\"{o}dinger equation of $H_{dw}$ accommodates one bound state with energy $\epsilon = 0$ 
and extended states of wave number $k$ with eigen energy $\epsilon_k = JSa^2 k^2 + 2AS$ \cite{winter}. 
\begin{eqnarray}
\psi_B(z) 
  &=& \sqrt{\frac{a}{2\Delta}} ~ \mbox{sech} \frac{z-q}{\Delta},   \\
\psi_k(z) 
  &=& \sqrt{\frac{a}{L}} ~ \frac{ ik\Delta - \tanh \frac{z-q}{\Delta} }{ \sqrt{1 + k^2 \Delta^2} } 
   ~ e^{ik(z-q)}.
\end{eqnarray}
$\psi_B(z)$ is the normal mode with the energy eigenvalue $\epsilon = 0$. 
This normal mode is bound within the potential well formed in the DW. 
Due to localization in the DW, this magnon mode may be called as the {\it bound magnon}. 
The wave function $\psi_B$ of the bound magnon is related to the DW structure function
$\theta_w$ by its spatial derivative $\partial_z \theta_w$, which implies a translation of the DW. 
This bound magnon is none other than the zero mode \cite{soliton}
of a DW, and tends to restore the translation symmetry of the FM nanowire. 
On the other hand, extended spin waves $\psi_k(z)$ of wave number $k$ 
($L$: length of FM wire) are plane waves with reduced amplitude 
in the DW, are characterized by the excitation energy gap and deforms the DW.

Local spin fields can be represented with the Holstein-Primakoff magnons:
$S_{i+} = \sqrt{2S-b_i^{\dag} b_i} ~b_i \simeq \sqrt{2S}b_i$, 
$S_{i-} = b_i^{\dag} \sqrt{2S-b_i^{\dag}b_i} \simeq \sqrt{2S} b_i^{\dag}$ and
$S_{iz} = S - b_i^{\dag} b_i$.
 With identification of the normal modes, we can define the corresponding magnon operators:
$b_k = \sum_i \psi_k^* (z_i) b_i$ (spin wave operators) and 
$b_w = \sum_i \psi_B^* (z_i) b_i$ (a bound magnon operator). 
Since $\psi_B$ and $\psi_k$'s form a complete set of orthonormal wave functions for $H_{dw}$, 
$b_w$ and $b_k$'s exhaust all possible normal modes of $H_0$ and 
the inverse relation can be readily written down,
\begin{eqnarray}
b_i &=& \sum_k \psi_k (z_i) b_k + \psi_B(z_i) b_w.
\end{eqnarray}
In terms of normal modes $b_w$ and $b_k$'s, $H_0$ is already diagonalized:
$H_0 = \sum_k \epsilon_k b_k^{\dag} b_k$.
Due to its zero energy, the bound magnon does not show up formally in $H_0$.

 Including the hard axis anisotropy (HAA), the magnon Hamiltonian can be diagonalized as $H_{mag} = H_B + H_{sw}$. 
The bound magnon still remains as a normal mode with energy $E_B = 0$. 
\begin{eqnarray}
H_B &=& - \frac{1}{2} KS \left(b_w - b_w^{\dag} \right)^2,   \\
\label{zeromode}
\begin{pmatrix} S_x \cr S_y \end{pmatrix} 
  &\propto& \psi_B (z) \begin{pmatrix} 1 \cr 0 \end{pmatrix}.
\end{eqnarray} 
Note that the bound magnon has no spin component along the hard axis, $S_{y} = 0$, 
but instead, its spin lies on the easy plane.
$S_x$ has the zero mode, which means that spins can rotate freely 
on the easy plane so that the DW can be shifted freely along the FM wire direction.
No zero mode in $S_y$ simply reflects no free rotation of spins away from the easy plane. 
On the other hand, the spin waves have spin excitations along two transverse directions 
with the increased excitation energy gap $E_k = \sqrt{\epsilon_k(\epsilon_k + 2KS)}$ under the HAA. 
\begin{eqnarray}
H_{sw} &=& \sum_k E_k a_k^{\dag} a_k,  \\
\label{spinwave}
\begin{pmatrix} S_x \cr S_y \end{pmatrix}
  &\propto& \psi_k (z) ~ \begin{pmatrix} u_x(k) \cr u_y(k) \end{pmatrix}.
\end{eqnarray}
Here $a_k$ is the spin wave boson operator under the HAA and a linear combination of $b_k$ and $b_{-k}^{\dag}$, 
and $u_{x/y}$ corresponds to the amplitudes of transverse spins.

 For the ordered local spins,  the spin texture can be described by the magnetization unit vectors, $\vec{m}_i$'s,
where $\vec{m}_i = \vec{m}(z_i)$ and $\vec{m} = \hat{z} \cos\theta + \hat{x} \sin\theta \cos\phi + \hat{y} \sin\theta \sin\phi$.
Under unitary transformation $u_i = \exp\left( - \frac{i}{2} \theta_i \hat{\phi}_i \cdot \vec{\sigma} \right)$,
which rotates the quantization axis of conduction electrons at site $i$ from the $z$ axis into $\vec{m}_i$
($c_{i\alpha} \mapsto d_{i\alpha} = u_{i\beta\alpha}^* c_{i\beta}$),
i.e., $u_i^{\dag} \vec{\sigma} \cdot \vec{m}_i u_i = \sigma^z$, 
the Hund coupling $H_{eS}$ is diagonalized and results in spin polarized conduction bands.
The kinetic term in $H_e$ introduces the current-spin coupling $H_{cS}$ or the Berry phase term\cite{bazaliy},
\begin{eqnarray}
\label{ham_cs1}
H_{cS} &=& \hbar v_s S  \sum_{i} (1 - \cos\theta_i ) \partial_z \phi_i.
\end{eqnarray} 
Here $v_s = -\frac{aI_s}{2Se}$ and $I_s$ is the spin current flowing in the system under electric field.
$I_s$ is computed from thermal average of $\hat{I}_{si} = \frac{et}{i\hbar} \sum_{\alpha} \alpha [ d_{i+1\alpha}^{\dag} d_{i\alpha} 
- d_{i\alpha}^{\dag} d_{i+1\alpha} ]$, which measures the spin polarized electric current from $i$ to $i+1$. 
$H_{cS}$ can be written in a compact form as
\begin{eqnarray}
\label{ham_cs}
H_{cS} &=& v_s \cal{P}, \\
\label{DW_P}
\cal{P} &=& \hbar S  \sum_{i} (1 - \cos\theta_i ) \partial_z \phi_i.
\end{eqnarray}
$\cal{P}$ is the DW linear momentum \cite{haldane} or the generator of DW translation as will be shown below.
Here angles are field variables.

 We now prove that $\cal{P}$ is the generator of DW translation or the linear momentum for DW.
For this purpose we consider the DW spin texture
$
|\Psi(\{z_i\}) > = \prod_i |S_i; \vec{m}_i> = \prod_i U_i (\vec{m}_i) |S_i; \hat{z}>,
$
where $U_i (\vec{m}_j) = \exp\left(-i\theta_j \hat{\phi}_j \cdot \vec{S}_i \right)$ rotates the orientation of $S_i$ from the $z$ axis 
into $\vec{m}_j$.
The DW state shifted to right by a lattice constant $a$ can be written as 
$|\Psi(\{z_i-a\})> = \prod_i |S_i; \vec{m}_{i-1}> = \prod_i U_i(\vec{m}_{i-1}) U_i^{\dag} (\vec{m}_i) |\Psi(\{z_i\})>$.
Writing $|\Psi(\{z_i-a\})> = \exp\left( - i\frac{a}{\hbar}\cal{P} \right) |\Psi(\{z_i\})>$, we can identify $\cal{P}$ as
\begin{eqnarray}
\cal{P} &=& \hbar \sum_i \vec{S}_i \cdot \left[  \vec{m}_i (1 - \cos\theta_i) \partial_z \phi_i - \vec{m}_i \times \partial_z \vec{m}_i  \right].
\end{eqnarray}
This quantum definition of $\cal{P}$ can also be obtained from Eq.~(\ref{DW_P}) by allowing small fluctuating angle or spin fields 
as in the normal mode expansion. 
Angles or $\vec{m}_i$ represent the DW solution. 
The first term is $c$-number ($\vec{S}_i \cdot \vec{m}_i = S_{iz} \simeq S$), while the second is the quantum correction $P$ and 
\begin{eqnarray}
\label{DWP_S}
P &=& -\hbar \sum_i \partial_z \theta_i ~ S_{iy},
\end{eqnarray}
for the transverse DW in our case.

 Under the finite spin current, $\theta_w(z)$ now becomes dynamical and thus,
the DW position $q$ is time-dependent. 
Two coupled Eqs. ~(\ref{snmodex}) and (\ref{snmodey}), under the current-spin coupling (\ref{ham_cs}),
are modified by two effects: dynamic $\theta_w$ and the spin current. 
In the rotating frame about the hard axis or the $y$ axis, the equation of motion for an operator $A$ is 
$i\hbar \partial_t A = [A, H_{\rm eff}]$, where the effective Hamiltonian $H_{\rm eff} = H_0 + H_1 + H_{cS} - \hbar \dot{\theta} S_y$ has 
an additional contribution from rotating angle $\theta$.
\begin{eqnarray}
\hbar \partial_t S_x &=& H_{dw} S_y + 2KS ~ S_y - \hbar S (v_s \partial_z \theta + \partial_t \theta),  \\
\hbar \partial_t S_y &=& - H_{dw} S_x. 
\end{eqnarray}
In general, the spin waves are coupled to the DW.
Normal modes under spin current are decoupled from DW  only when $\frac{dq}{dt} = v_s$,
i.e., the DW moves with the \textit{adiabatic} velocity $v_s$.
If the DW absorbs with full efficiency the spin angular momentum transferred from the spin current, 
there will be no Doppler shift \cite{bazaliy,tatara,li_zhang,Macd} in the spin wave energy spectrum. 
If not, spin angular momentum from the spin current will be transferred to exciting spin waves.

 The DW dynamics is determined by the bound magnon Hamiltonian, $H_B$ and $H_{cS}$. 
Since $\partial_z \theta \propto \psi_B$ is finite only near the DW, 
the main contribution to $P$ comes from spins
in the DW and the number of contributing spins is roughly $\Delta/a$.  
Furthermore, owing to $H_{dw} \sin\theta_w =0$, we have the identity $[P, H_{mag}] = 0$ 
such that $P$ is the constant of motion. 
This is a simple mathematical manifestation of translational symmetry for a DW in an infinite FM nanowire.  
$P$ can be represented in terms of the bound magnon as
\begin{eqnarray}
\label{DWP_B}
P &=& -i\hbar \sqrt{\frac{S}{a\Delta}} ( b_w^{\dag} - b_w ).
\end{eqnarray}
$H_B$ can be interpreted as the kinetic Hamiltonian of the bound magnon by noting that $H_B$ can be written in terms of $P$ as
\begin{eqnarray}
\label{ham_bm}
H_B &=& \frac{K \Delta a}{2\hbar^2} P^2 
  = \frac{P^2}{2M_{dw}},
\end{eqnarray}
where the bound magnon mass is defined as $M_{dw} \equiv \frac{\hbar^2}{K \Delta a}$.  
The DW or D\"{o}ring mass defined in the classical approach \cite{doring} is none other than 
the mass of the bound magnon or the zero mode in a ferromagnetic DW.

 The effect of the hard axis anisotropy $K$ is threefold. For extended spin waves, their energy
gap is enhanced such that they become much harder to excite. 
The hard axis anisotropy confines the bound magnon to have spin components only 
on the easy plane, but no component along the hard axis. 
The bound magnon acquires its inertia due to the hard axis anisotropy.

 Dropping $c$-number from $\cal{P}$, the current-spin coupling Eq.~(\ref{ham_cs}) becomes 
\begin{eqnarray}
\label{ham_cs2}
H_{cS} &=& v_s P, 
\end{eqnarray}
which is the same for both $\phi_w = 0, \pi$.  Note that $P (\phi_w=\pi) = - P(\phi_w = 0)$. 
Under spin currents, the system retains a translational symmetry.
The action of $H_{cS}$ on the DW can be most easily understood in terms of the Schr\"{o}dinger equation,
\begin{eqnarray}
i\hbar \partial_t |\Psi (\{z_i\},t) > &=& (H_B + H_{cS}) |\Psi(\{z_i\},t)>.
\end{eqnarray} 
Denoting the DW state as $|\Psi(\{z_i\})>$ when $v_s = 0$, we find that $|\Psi(\{z_i\},t)> = \exp\left(-iv_st P/\hbar \right) |\Psi(\{z_i\})>
 = |\Psi(\{z_i-v_st\})>$.  The DW motion with velocity $v_s$ is induced by the spin current.

 The intrinsic pinning was claimed \cite{tatara} to be induced by HAA 
of a perfect FM nanowire. Spin currents rotate the local spins away from the easy plane 
and the HAA field acts as a blockade \cite{tatara} to the DW motion 
and generates the intrinsic pinning. 
The quantum approach clearly shows the absence of the intrinsic pinning
and is consistent with the translation symmetry of DWs under spin currents. 
The DW absorbs the spin angular momentum from the spin current via the bound magnon, and thereby
avoids the tilting of local spins away from the easy plane.

There is (no) translation symmetry for DW in FM wires under spin currents (magnetic fields). 
The energy damping torque like the Gilbert type is prerequisite for the steady domain wall motion 
under magnetic fields. Energy dissipation via damping uses up the Zeeman energy and sets the DW in motion.
On the other hand, the DW under spin currents absorbs spin angular momentum via the bound magnon from 
conduction electrons and can move even without damping.

  Our theory is based on the perfect FM nanowires without spin damping. 
With the fully efficient absorption of spin angular momentum, 
the spin current sets the DW in motion with the adiabatic velocity $v_s$,
the spin waves are decoupled from the DW motion, and no Doppler effect is expected in the spin wave energy spectrum.
Note that the spin wave energy shift under spin currents was observed \cite{doppler} in FM nanowires with {\it uniform} magnetization.
According to the phenomenological LLG equation \cite{barnes,li_zhang,thiaville}, 
the DW velocity is modified from the adiabatic value $v_s$ by the so-called $\alpha$ \cite{gilbert} and $\beta$ \cite{li_zhang} damping torques.
In this case more careful study \cite{jvkim} is required for elucidation of the Doppler effect in the spin wave energy spectrum.


 In summary we studied the domain wall motion under spin current, based on the $s$-$d$ model Hamiltonian. 
We found that the bound magnon plays an important role in the domain wall dynamics. 
Since the bound magnon is localized to the domain wall and has zero excitation energy,   
a spin current, without energy cost, transfers spin angular momentum to the domain wall via the bound magnon.
Furthermore the hard axis anisotropy confines the bound magnon on the easy plane 
so that local spins can absorb spin angular momentum from the spin current 
and rotate about the hard axis without tilting away from easy plane.
The bound magnon acquires its inertia due to the hard axis anisotropy and remains a free particle even under spin currents.
With the full transfer of spin angular momentum from the spin current to the DW, 
spin waves are decoupled from the DW motion and no Doppler shift is expected for the spin waves. 
The bound magnon-mediated spin transfer mechanism leads to the free motion of bound magnon 
or domain wall under spin currents.

This work was supported in part by a Grant-in-Aid for Scientific Research in Priority Area `Creation and control of spin current' from
the Ministry of Education, Culture, Sports, Science and Technology
(MEXT), Japan, a Grant-in-Aid for Scientific Research (A) from MEXT,
Japan, the Next Generation Supercomputing Project of MEXT, Japan, and
in part by the Korea Science and Engineering Foundation (KOSEF) grant
funded by the Korea government (MOST) (No. R01-2005-000-10303-0). We
thank S. E. Barnes and Chanju Kim for stimulating discussions.

\end{document}